# Simplicial models of social aggregation I

Mirco A. Mannucci, Lisa Sparks, Daniele C. Struppa

This article is dedicated to the memory of Jonathan M. Beck (d. 2006), who taught one of us to search for the hidden, endless treasures of the simplicial ladder.

Abstract

This paper presents the foundational ideas for a new way of modeling social aggregation. Traditional approaches have been using network theory, and the theory of random networks. Under that paradigm, every social agent is represented by a node, and every social interaction is represented by a segment connecting two nodes. Early work in family interactions, as well as more recent work in the study of terrorist organizations, shows that network modeling may be insufficient to describe the complexity of human social structures. Specifically, network theory does not seem to have enough flexibility to represent higher order aggregations, where several agents interact as a group, rather than as a collection of pairs. The model we present here uses a well established mathematical theory, the theory of simplicial complexes, to address this complex issue prevalent in interpersonal and intergroup communication. The theory enables us to provide a richer graphical representation of social interactions, and to determine quantitative mechanisms to describe the robustness of a social structure. We also propose a methodology to create random simplicial complexes, with the purpose of providing a new method to simulate computationally the creation and disgregation of social structures. Finally, we propose several measures which could be taken and observed in order to describe and study an actual social aggregation occurring in interpersonal and intergroup contexts.

Toward a dynamic representation of interpersonal and intergroup communication:
Simplicial models of social aggregation

A topic of great interest in modern social sciences is the modeling of social aggregation in a variety of forms. Social scientists are interested in the theory of group formation, group evolution, and are looking for a variety of models which could help them to both describe group static and its dynamic evolution. Scholars have been particularly interested in examining group dynamics and decision-making in family, and health risk communication contexts, particularly when it comes to the potential to offer tools to act on groups in terms of either strengthening or weakening them (see e.g., Ayres, 1983; Fisher, 1953; James, 1953; Harwood & Sparks, 2003; Noller & Fitzpatrick, 1990; Nussbaum, 1989; Sillars & Wilmot, 1989; Sparks, 2003; Sparks, 2005; Sparks & Harwood, in press; Sparks, Kreps, Botan, & Rowan, 2005; Sparks-Bethea, 2002; Wheeler & Nezlek, 1977).

This paper proposes a new theoretical model that offers a dynamic representation of interpersonal and intergroup communication by describing social aggregations, providing a quantitative analysis of these social groups, and offering the possibility of computer simulations of the evolution of such groups. Although various interpersonal and intergroup contexts certainly possess unique characteristics (see e.g., Gallois & Giles, 1998), the simplicial model of social aggregation provides an important mathematically based dynamic representation that is grounded in prior theoretical work in interpersonal communication and intergroup relations.

An example which has regretfully become quite topical is the study of terrorist networks, which only recently has been a topic of interest for communication scholars (see e.g., Sparks, 2005; Sparks, Kreps, Botan, & Rowan, 2005). The assumption here is that one can model a terrorist group (or a coalition of groups) via networks. In this model, every terrorist is represented by a node in the network, and every time two terrorists are interacting with each other, this is represented by an edge connecting the two nodes. This model is interesting because mathematicians have developed a reliable and deep theory of networks, and because one can add detail to the networks by allowing, for example, multiple edges across nodes, as well as directed edges, colored edges, and finally weighted edges. Thus, the quality of the simulation becomes increasingly good. The purpose of modeling terrorist structures with networks is to provide our intelligence agencies with visualization models, but also with tools to predict the evolution of the network. On this point, the foremost theory is the so-called Random Graph Theory (RGT) introduced by Erdos and Renyi (1960). Recently RGT has been partly superseded by more realistic models, such as the theory of small-worlds (Watts 1999a, 1999b, 2004), or the theory of scale-free networks (Barabasi, 2003; Barabasi et al., 2002): an excellent and comprehensive survey is given by Dorogotsev and Mendes (2001). Of related interest is the study of the topological properties of networks with the purpose of classifying them, but also of finding ways to destroy or weaken such networks.

   However, network theory as a way to discuss social aggregation is not limited to military applications. As Sparks has pointed out in (Sparks-Bethea, 2002), graph and network theory are used in the study of interpersonal communication as a tool to describe, model, negotiate, re-negotiate, and control situations. Further, researchers of relationships are frequently concerned with understanding marriage, romance, or friendship; yet often forget about the matrix of associations in which individuals and any one of their relationships are embedded. As Bateson (1984) suggested, to attempt to understand an individual or any one of their relationships separate from their social matrix is misleading. Although dyadic communication is essential to an understanding of relationships (see e.g., Fisher, 1953; James, 1953; Wheeler & Nezlek, 1977), it is equally essential that researchers of relationships gain a broader understanding of the dyad as it is intertwined in other relationships. Thus, while interpersonal communication is grounded in interactive processes, behaviors, symbolic exchanges, shared meaning, social cognition, context, consciousness and intent, individual differences, and the like, intergroup communication also takes an individual's social categories into account. Intergroup communication is most often defined as communication behaviors exhibited by one or more individuals toward one or more other individuals that is based on the individuals' identification of themselves and others as belonging to different social categories (see Harwood, Giles, & Ryan, 1995; Hajek & Giles, 2002). Thus, it is important to understand the essential links between interpersonal communication and intergroup communication. It is this broader understanding that we hope to be able to offer with our model. To offer a specific example, the literature on family relationships and family decision-making suggests that when the fundamental dyad (husband and wife) is supplemented by the arrival, for example, of an aging parent the core relationship undergoes a resilient modification of communicative behaviors to adjust to the new situation (see Sparks-Bethea, 2002). The dyad becomes then a triangle, and the effects can be quite unexpected. This particular example, however, is important because it shows in which sense one may say that networks are inadequate as a tool to understand social aggregation. Let us consider, again, the husband-wife dyad. Its natural network representation is a graph with two nodes (H for husband, and W for wife), and one edge connecting them, which represent the relationship and the communication between them. When a third person enters this dyad, for example an elder parent, the network representation allows only three different representations. In all cases we will have a graph with three nodes (H, W, and P for parent), and with an edge HW between husband and wife. However, we can represent the actual communicative structure by either having an edge HP between H and P and an edge WP between W and P or not. So, three cases arise: HP and WP are both present, or HP is present and WP is not, or finally WP is present but HP is not.

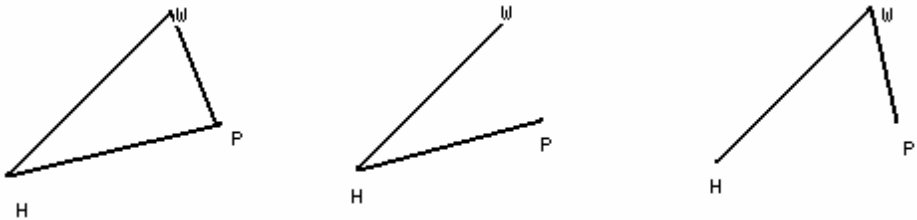

The last two of these graphs are isomorphic, in the sense that they are simply connected (i.e. they will become disconnected if node H or W are removed). The first graph, however, has three nodes and three edges, and it cannot be disconnected by eliminating one of the nodes (for example if one of the family members dies). For this reason, this is a model of a stronger group. Nevertheless, social scientists have discovered (see e.g., Sparks-Bethea, 2002) that even when the model for the triad is the triangle, there are very distinctive social situations which cannot be well modeled by the triangle shaped network. In particular, we can discuss at least two different situations. In one, the three members of the family talk to each other, on a one-to-one basis, but don't really create a unified family. W talks to H, and both of them talk to P, but there is little interaction of the three as a family. There is no family dinner, so to speak. As it turns out, this situation may have really negative effects on the strength of the connection between H and W (as it could be indicated by a weakening of the weight of the edge HW). A very different situation takes place when in addition to the three dyadic interactions, there is also an interaction at the group level. This is the situation in which the three pairs have actually merged into a family. As it is clear, networks cannot directly model such a case, though at least some of the situations could be handled through the development of dynamical networks with conditional probabilities

Once again, Sparks noted in (Sparks-Bethea, 2002) that the study of triadic relationships offers an opportunity for researchers to move beyond the dyad as a stand-alone functioning unit. Wilmot and Sillars (1989) point out that "Just as individual responses fail to reveal everything of interest about interpersonal relationships, dyadic relationships fail to reveal everything of interests about a person's network of relationships" (p.128). Studying the fluid formation of triadic relationships may add a much-needed element to better understand the needs of the functioning marital dyad as it is influenced by other relationships.

Communication scholars have generally neglected the study of "triadic" relationships. Exceptions exist in the health communication domain (e.g., Beisecker, 1989), the family communication domain (e.g., Long & Mancini, 1990; McHale, 1995; Wilmot, 1987; Sparks-Bethea, 2002), and the personal relationship domain (Klein & Milardo, 1993). These studies, however, have failed to systematically examine triadic relationships, particularly as they change and adapt across the life span. What may be important, at least for individuals in these triadic relationships, is to understand the extent to which individuals are influencing each other in these connected relationships.

If we go back to the model we were discussing concerning terrorist networks, we see that also in that case the network description is insufficient. This is why, in everyday language, we hear talk of 'terror cells', a notion which indicates a small group of terrorists, who are in constant communication with each other and whose level of cohesion (cognitive, ideological, organizational) is higher than the one offered by a network model.

We need, therefore, a different model, which includes networks, but that also allows for higher order groups, in which three or more people give rise to something more than a set of dyadic conversations. Fortunately, such an object is well known in mathematics, and in fact some would even go as far as claiming (see for instance Beck, 1979) that it is one of the fundamental building blocks of mathematics. We are referring here to what is called "Abstract Simplicial Complex" and, which can be described (see next section for a detailed and formally correct definition) as a higher dimensional version of a network. When we look at isolated nodes, we have a 0-dimensional structure. When we connect some (or all) of the nodes, we obtain a structure which is 1-dimensional (though it may still have 0-dimensional components). In simplicial terminology, nodes are known as 0-simplexes, while edges are known as 1-simplexes. When we connect nodes with edges we have a simplicial complex. Moving to higher dimension, a triangle is known as a 2-simplex. Note (very important point for what follows), that a triangle (a 2-simplex) has three faces (we could call them sides) which are segments, i.e. 1-simplexes. In turn, every 1-simplex has two faces (we could call them border points) which are 0-simplexes. We can therefore think that our 2-simplexes must be faces of something else we should call a 3-simplex. This is exactly the case, as 3-simplexes are defined as tetrahedrons (i.e. pyramids with three faces, each of which is a triangle).

Before we offer a few examples, and some formal definitions, we wish to point out that the idea of using simplexes to describe and model social interactions is not totally new, as it was developed by R. Atkin in his theory of Q-analysis (Atkin 1972, 1974, 1977) as well as in Legrand's recent work (Legrand, 2002). However, as the reader will see, there are many fundamental differences between our approach and Atkin's. To begin with, Atkin uses simplicial complexes to represent relations in a Cartesian product of two sets (i.e. binary relations). In our point of view, simplicial complexes indicate actual social entities, whereas relations (binary and n-ary alike) that comprise them are, literally, their faces. Just as a human body is more than a mere aggregate of its organs, social groups are more than their constituents. Even more crucial is the fact that Atkin's model is essentially static. We introduce, on the other hand, a probabilistic, dynamic approach to the evolution of such simplexes, along the lines drawn by Erdos in his approach to random graphs. Social groups (and the simplicial objects representing them), are to be thought of as living, evolving, beings. Dynamics is, here, at the very core of the theory.

We will give the details of the construction we propose in the next section, but for now we want to point out that a family triad as the one illustrated before can now be modeled in four ways:

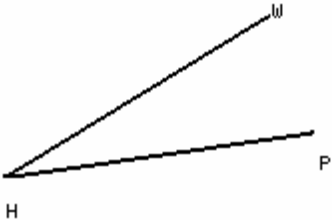

Nodes {H,W,P}; Edges {HW,HP}

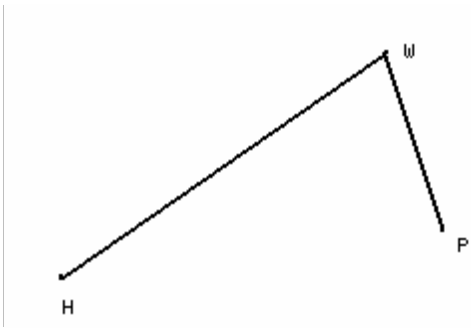

Nodes {H,W,P}; Edges {HW,WP}

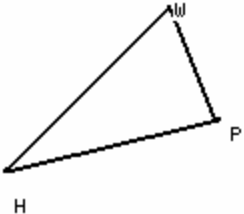

Nodes {H,W,P}; Edges {HW,HP,WP}

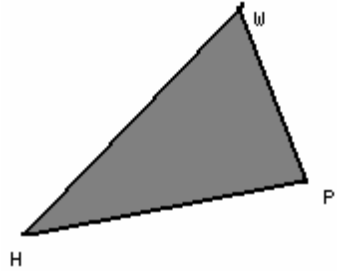

Nodes {H,W,P}; Edges {HW,HP,WP}; Triangles {HWP}.

The four representations are four different simplicial complexes. The first three are one-dimensional with three 0-simplexes (the nodes) and either two or three 1-simplexes (the edges). The last representation is the only two dimensional simplicial complex, with three 0-simplexes, three 1-simplexes, and one 2-simplex (the triangle which represents the family relationship). It is clear, even without developing a full theory of simplicial models, that we now have at our disposal a more complex and satisfactory way to describe the possible outcomes of the arrival of a parent in a dyadic family.

In this paper, we want to offer the theory of simplexes as a new and robust model to describe social aggregation. We intend this to be the first in a series of papers where we approach several important issues, and we want to use this work to set the foundation for our future research. To this purpose, section 2 will formalize and recall some notions from the theory of simplicial complexes. Section 3 will discuss what could be called random simplicial topology, namely a model of how we can generate random simplicial complexes, in an attempt to offer a model for group formation and evolution. Finally, Section 4 will begin a discussion on how these models can be implemented in a computational fashion, which will make them useful for social scientists not only to understand and visualize social aggregations, but also to offer theoretical tools to help strengthen or weaken these aggregations.

An introduction to simplicial complexes

This section is devoted to a brief introduction to the theory of simplicial complexes. This is a rather complex theory, which is very well established, and the reader interested in more detail is referred to several standard references in algebraic and combinatorial topology such as (Munkres, 1984). In this section, however, we will only provide those notions which are necessary for the development of our models, and we will recall some of the fundamental results, of course without proving them. The next few paragraphs may appear a bit technical, and in particular the reader may be baffled as to why we are using higher dimensional spaces (instead of limiting ourselves to the well known three-dimensional space). The reason for that will be apparent once we start introducing some concrete examples. We are asking, therefore, patience while reading the next few pages.

The notion of simplex is a generalization to higher dimension of the intuitive notions of triangles and pyramids. But let us proceed in order. A simplex of dimension zero is a point P. A simplex of dimension one is a segment, including its end points. If P and Q are the end points, we denote by [P,Q] this one-simplex. A simplex of dimension two, or two-simplex, is the inside of an equilateral triangle [P,Q,R] whose vertices are the points P, Q, and R. As for the 1-simplex, we ask that the triangle itself is part of the 2-simplex. A 3-simplex (or a simplex of dimension three) [P,Q,R,S] is the inside of a tetrahedron (the fancy mathematical name for the pyramid with four faces, each of which is an equilateral triangle), again including the faces. The next steps are a bit harder to visualize as they use higher dimensions. We will introduce the simplex of dimension four, and then we will give the general definition of a simplex in dimension n. To construct the simplex in dimension four, we start with a regular tetrahedron [P,Q,R,S] (the 3-simplex) and then we take a point T in the fourth dimension, such that the distance (in the four-dimensional space) between T and each of the points P,Q,R, and S is the same. We then fill up the object [P,Q,R,S,T] in the four-space by taking all the points in the four space which are "inside" the boundary created by the points P,Q,R,S, and T (in the same way in which a triangle is the set of points "inside" three given points P,Q, and

R. The resulting object is a 4-simplex or pentatope. Mathematicians have an expression to indicate this filling-up process, namely they say that the pentatope is the convex hull of P,Q,R,S, and T. Let us describe what convex hull is, since this notion will be used in the sequel. Let us start with two points P and Q. The convex hull of these two points is the segment between them. If we now take three points P, Q, and R, the convex hull of them is obtained by taking the convex hull of each pair, and then the convex hull of all the pairs of points which have been obtained through this procedure, and so on, until there is nothing else left to do. So, if three points are given, the first step is to take the convex hull of P and Q, i.e. the segment [P,Q], then the convex hull of Q and R, i.e. the segment [Q,R], and then the convex hull of P and R, i.e. the segment [P,R]. We have obtained the boundary of the triangle with vertices P, Q, and R. We now continue the process by constructing the convex hull of each pair of points taken from the boundary. So, for example, we take a point on [P,Q] and a point on [Q,R] and we build their convex hull. This will give us a segment which falls "inside" the boundary. Once we are finished with this process, we obtain the entire triangle [P,Q,R]. At that point, even if we try to build more convex hulls, we don't find any new points, and the process is actually over. The triangle [P,Q,R] is the convex hull of the three points P,Q, and R. Similarly, the 3-simplex [P,Q,R,S], i.e. the tetrahedron, is the convex hull of the points P, Q, R, and S. The road to generalization is now obvious. In dimension n, the n-simplex is the convex hull of n+1 points $P_0, P_1, P_2, \ldots, P_n$. For the sake of completeness, we need to mention that the points need to be chosen in such a way that the directions $P_1-P_0, \ldots, P_n-P_0$ are all different.

We should mention that this way of describing simplexes can be at once simplified and generalized if one has enough language from the theory of categories, and the definition becomes a bit quicker in that case. The mathematically oriented reader is referred to the classic book (May, 1993) for more details. We wish to point out that using the categorical machinery it is possible to define simplicial objects in environments that have more structure than just being a set. This is important, because it allows us to model groups wherein individuals and their relations (dyadic and polyadic) carry additional pieces of information (for instance, weights, labels, different types, etc.). This "categorical" approach will be pursued by the authors in a more abstract and mathematically oriented work (Anonymous, 2006).

The objects we are ultimately interested in, however, are not simplexes, but rather simplicial complexes. To define what a simplicial complex is, we need to introduce the notion of "face" of a simplex. This is the natural generalization of the notion of "face" of a polyhedron. So, the faces of a 1-simplex are its two end points. The faces of a 2-simplex (a triangle) are its three sides, the faces of a 3-simplex (a pyramid) are its four faces (each of which is a triangle), the faces of a 4-simplex (a pentatope) [P,Q,R,S,T] are the five 3-simplexes which one obtain from the pentatope by ignoring one vertex at a time (i.e. they are [P,Q,R,S]. [P,Q,R,T], [P,Q,S,T], [P,R,S,T], and [Q,R,S,T]), and so on and so forth. Note that the face of an n-simplex is always an (n-1)-simplex.

Given a simplex, we can consider its faces, and then the faces of its faces, and so on. This leads us to the notion of m-face of a simplex. Quite simply, given an n-simplex $S=[P_0,P_1,…,P_n]$, we say that the points $P_j$ are the 0-faces of S, we say that the 1-simplexes $[P_i,P_j]$ are the 1-faces, the 2-simplexes $[P_i,P_j,P_k]$ are the 2-faces, etc.

We can now define a simplicial complex. By this word we mean the union of several simplexes (possibly of different dimensions) such that if the intersection of two simplexes is not empty, then the intersection itself is an m-face for both simplexes. Here below we have a picture of a simplicial complex, and a pictures of an object which, while the union of simplexes, is NOT a simplicial complex.

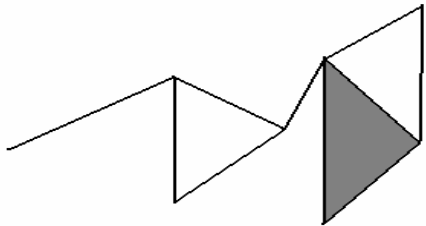

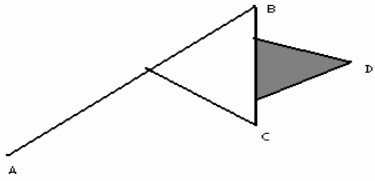

Before we introduce some of the computational tools which allow us to describe properties of simplicial complexes, we would like to offer some examples which explain why higher dimensions are needed to use simplicial complexes as models for social aggregation. What follows are simply examples to show how simplicial complexes can be used to visualize such aggregations. No computations of any type are attempted.

*Example 1*. A family is composed of dad, mom, and a child. This is a strong family, in which the three components actually come together and work as a unit. The simplex which describes this family is a 2-simplex. The nodes are D, M, C (dad, mom, and child); the 1-faces describe the interaction between any two of them (D and M, or D and C, etc.), and the entire simplex describe the interaction of the three components as a family. The graphical representation is indicated below.

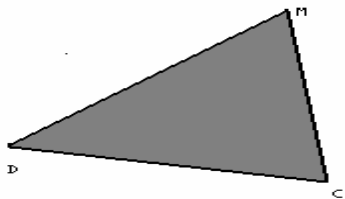

*Example 2*. Suppose now the family splits as the consequence of a divorce. The new structure is a simplicial complex of dimension 1. It is made of three nodes (as before), and three 1-simplexes (the relationship between mom and child, the relationship between dad and child, and the relationship – albeit functioning differently than when married and most often functioning negatively post-divorce – between mom and dad). What is now missing is the 2-face, because the three components do not come together as an intact family any more. The representation is below.

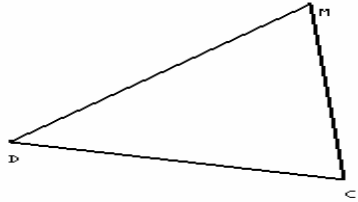

*Example 3.* Mom now marries again, and the child lives mostly with mom and her new stepdad. The new family M, C, S (for stepdad) is a strong one, which operates as a family. The representation now is a simplicial complex in which we have simplexes of different dimensions. There are four nodes (D, M, C, S), and two triangles. One is the triangle representing the original (now broken) family, and it involves nodes D, M, and C. The dimension of this triangle is one, as only the edges exist. The second triangle represents the new family, and it involves nodes M, C, and S. Because this is a strong family, the representation is actually a 2-simplex (as we had originally in example 1). Nodes S and D are not connected by an edge (they are the only ones in this picture not to have a connection). So we have one simplex of dimension 2 (the family [M,C,S]), to which we attach two simplexes of dimension 1 ([M,D] and [C,D]). These two 1-simplexes are in turn attached on the 0-simplex D. See below for the picture describing this situation.

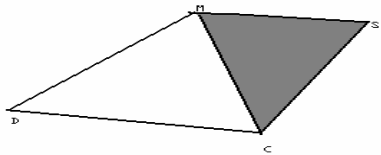

*Example 4.* A variation on this example would occur if mom and dad reach a point in which their relationship is so strained that they don't even communicate any more. This would destroy the edge connecting M and D, and the resulting picture would be as follows:

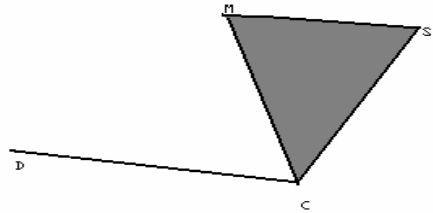

*Example 5.* Building on example 3, let us now assume that the new family [M,C,S] has a new baby B. The new baby is fully accepted by C, and the family now must be represented as a 3-simplex [M,C,S,B]. On this simplex we have four faces, each of which describes the interaction of the subfamily which one has when one of the components is absent (when stepdad is away for work, [M,C,B] still work together as a smaller family, and similarly, when C is away for visitation with her dad, [M,S,B] are a fully functioning family, etc.). The new baby B and the original dad D are not connected, and so the simplicial complex we need to represent this more complex family is now a tetrahedron (representing [M,C,S,B]) and two 1-simplexes (the relations [M,D], and [C,D]). See below for a graphic representation.

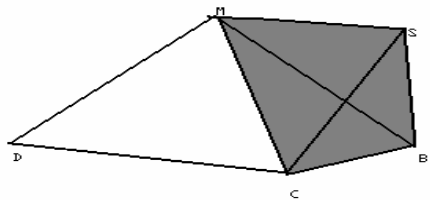

*Example 6.* Finally, a second little baby is born in this new family, let's call her A. The representation now requires us to use a pentatope [M,C,S,B,A] to describe the new family, plus the old 1-simplexes. This last example, which we cannot graphically represent, explains why we need higher dimensional simplexes when we want to describe increasingly complex social structures.

The few examples we have given above indicate how this mathematical tool can be used in a dynamic way to represent social aggregates, and their change through time. We can think of a social structure as being a function F from the set of positive numbers with values in the space of simplicial complexes[1]. F(0) represent the initial state of a social structure, and as times t increases, F(t) represents the evolution of that social structure. As new relationships are born, the complexity of F(t) will increase, but at the same time F(t) can become simpler if some of the members of the social structure die (or leave the structure in some other less dramatic way), or if relationships are broken (this reduces the dimensionality of the simplicial complex, at least in some of its components). Thus, simplicial complexes can easily be seen as helpful to the social scientist for at least the following three applications:

1. Simplicial complexes can be used to model and visualize a social structure.
2. Computational tools from the theory of simplicial complexes can be used to describe and modify a given social structure.
3. A dynamical theory of simplicial complexes can be used to simulate the growth and structure of a given organization, with known initial conditions.

This current paper, as the first one in which the theory of simplicial complexes is applied to social aggregation, focuses essentially on 1. Nevertheless, we want to give some ideas of how 2 and 3 could be possible. Specifically, in this section we want to discuss some of the computational tools which are available to describe simplicial complexes. In the next section we will discuss how to set up a dynamical theory of simplicial complexes. Both ideas will be then expanded in future papers. Let us start with some very simple examples, and then we will try to gain some theoretical ideas from them.

---

[1] In a subsequent work (Mannucci, Sparks, Struppa, 2006b) we will consider more general frameworks for social dynamics. For the mathematically savvy reader, we say that we shall employ sheaf models from a small category ( generalized "time") with values in the category of simplicial sets (or even simplicial objects in some ambient universe).

*Example 7.* Consider simplex S made of two 0-simplexes [P] and [Q]. It represents two people who do not interact with each other. Two strangers. Now consider simplex T=[P,Q], which represents those two same people who now interact (maybe as friends, maybe as enemies, maybe as lovers…). The question is: how do we distinguish (from a mathematical point of view) these two simplexes? There are at least two ways of doing this. One is to note that S has dimension zero (its components are all 0-simplexes), while T is of dimension one. This teaches us that "dimension" is a numerical concept which can be used to describe different simplexes or simplicial complexes. Another way to distinguish S from T is to note that "S is made of two disconnected pieces", while T is just one piece (I can navigate along T, without ever lifting the pencil, which I clearly cannot do on S). We say that T is connected, and that S is not. Or, we say that the 0-th Betti number of S is two (S is made of two pieces) while the 0-th Betti number of T is one (T is made of a single piece). The notion of connectedness, by itself, is not too useful, as it would not distinguish a simplicial complex made of three unconnected nodes from one made of two unconnected nodes. More relevant is the Betti number (the name comes from a famous Italian mathematician of the end of the nineteenth century), which tells us how many disjoint connected components make up the simplicial complex.

*Example 8.* Unfortunately, the 0-th Betti number is not enough to characterize a simplicial complex. For example the complex S={[P],[Q]} and the complex T={[P,Q],[R,S]} both have 0-th Betti numbers equal to two. Complex S is made of two disconnected points. Complex T is made of two disconnected segments. Either way, they are made of two disjoint components, and so they have the same 0-th Betti number. What would diversify them, on the other hand, would be to note that S has no components of dimension one. To do so, we introduce the 1-st Betti number, which counts exactly how many "holes" of dimension one exist. In this way, we see that S's 1-st Betti number is zero, while T's first Betti number is two. Thus, we have been able to differentiate these two simplicial complexes, through the use of two Betti numbers. If we use the standard notation $b_0$ and $b_1$ to denote the Betti numbers, we see that $b_0(S)=b_0(T)=2$, but $b_1(S)=0$, while $b_1(T)=2$.

The definition of Betti numbers can be extended to higher dimensions, and this is necessary to differentiate examples which otherwise would be undistinguishable. Regretfully, the computation of these higher order Betti numbers (as well as their precise definition) ends up being a bit more complicated than what appears from the previous two examples. For example, if we consider the simplicial complex T of example 1 (the triadic family before the divorce), we have $b_0(T)=1$, $b_1(T)=0$, and $b_2(T)=1$. On the other hand, if we consider the simplicial complex S of example 2 (a triadic family after a divorce), we see that $b_0(S)=1$, $b_1(S)=1$, and $b_2(S)=0$ (we could say that a one-dimensional hole in the social tissue has been created).

But what is important from our point of view, is that a general theorem of simplicial topology says that if two complexes have the same Betti numbers for every dimension, then the two complexes are equivalent, in some suitable sense (they will have the same degree of connectivity in each dimension). Equivalent complexes, from our point of view, would represent social aggregations with the same degree of cohesiveness. Unfortunately homological/homotopical techniques only detect differences up to so-called homotopy equivalence (i.e. two objects are considered the same if they can be deformed into one another without breaking them), which is weaker than topological isomorphism (here objects are considered indistinguishable if all of their topological properties are indeed the same). Therefore it may be useful, in certain context, to consider the finer notion of topological isomorphism to distinguish between social groups. For instance, a couple and a single are homotopically equivalent (neither has any "holes" and the single is a "retract" of the couple), but they are clearly very different social groups (their topology is indeed quite different: the couple can break down, leaving two isolated points).

The reader who is interested in more detail in the theory of Betti numbers (and the related notions of homology and cohomology) is referred for example to standard references in algebraic topology, such as (Munkres 1984).

Random simplicial complexes

In this section we want to set the stage for a theory designed to model the evolution of simplicial complexes, and as such designed to model the evolution of social aggregates. The ideas described in this section are novel and due to the authors. However, we acknowledge the influence of the early fundamental work of Erdos and Renyi on random graphs (Erdos & Renyi, 1960; Bollobas, 2001) The work of Erdos and Renyi was extremely influential, not just within the mathematical community, but also because it spawned a series of important works on evolving networks, beginning with (Watts 1999a, 1999b). It was the availability of these tools, we believe, that made it so appealing to try to use network theory to describe and discuss terrorist groups. Unfortunately, as we have pointed out, graph models seem insufficient, and we hope that the ideas contained in this section will offer a more powerful and accurate model.

Let us begin by recalling the simple model of Erdos and Renyi. They consider a graph with undirected edges (i.e. a simplicial complex of dimension at most one), and they make the following basic assumptions:

(i) the total number of nodes in the graph is fixed, N;

(ii) the probability that two arbitrary nodes are connected equals a fixed number p. It is obvious to see that, on the average, the graph will contain pN(N-1)/2 edges, and in fact Erdos and Renyi go on to compute other important numerical characteristics of these graphs. It is worth emphasizing that, despite the obvious simplicity of this structure, random graphs actually turn out to be quite applicable, and a very interesting mathematical object. It is also clear that more complicated objects can be obtained, for example, if one allows the probability p not to be a constant on the graph, or if one allows the number N to increase as time increases. We are now ready to introduce our first new concept.

*Definition 1*. A random simplicial complex S(n,N,p) of size n, dimension at most N, and constant probability p is constructed as follows. First we fix a number n of points, which are the 0-simplexes of the complex. We call $S_0$ the set of these numbers. Given two vertices, the likelihood that they are connected by an edge is p. A realization of this random process yields the 1-skeleton of the complex, which we call $S_1$. Given three vertices A, B, and C such that they are all connected to each other, the likelihood that [A,B,C] is a simplex in the complex is p. The realization of this process now yields the 2-skeleton of the complex, denoted by $S_2$. We now continue this construction up to simplexes of dimension N. We set $S=S_N$.

Note that we use an abuse of language, since we call a random simplicial complex what is actually a realization of a random process. We will be careful to distinguish the two when and if the distinction becomes relevant. For now, let us give a few examples.

*Example 1.* A random simplicial complex S(n,1,p) is nothing but a random graph according to Erdos and Renyi.

*Example 2.* Let us consider S(3,2,0.5), and let us study the possible realizations of this random simplicial complex. First we note we start with three points A, B, C. The likelihood that they are connected (two at a time) is 0.5, which means the following realizations can occur at first stage (all of them with likelihood 0.125):

$S_0$ = {A, B, C} (none of the possible edges exists);

$S_1$ = {A, B, C, [A.B]}; $S_1$ = {A, B, C, [A,C]}; $S_1$ = {A, B, C, [B,C]} ;

$S_1$ = {A, B, C, [A.B], [A,C]}; $S_1$ = {A, B, C, [A.B], [B,C]};

$S_1$ = {A, B, C, [A.C], [B,C]); and $S_1$ = {A, B, C, [A.B],[A,C],[B,C]}.

In all cases, except the last one, the simplicial complex is complete, as no 2-simplexes are possible, but the last case presents the possibility of creating a 2-simplex or not, again with probability 0.5. Thus, the possible realizations of the complex are, in fact, the first seven listed above (with probability 0.125) and then two more realizations (each with probability 0.0625), namely $S_1$ = {A, B, C, [A.B],[A,C],[B,C]}, and $S_2$ = {A, B, C, [A.B],[A,C],[B,C]}.

It is clear how quickly complex this procedure can become. The point here is that we have a mechanism to create realizations of simplicial complexes, and that we can now envision situations in which we can put in motion such a mechanism, to describe evolving random simplicial complexes. This is the definition we are aiming for:

*Definition 2*. Let n(t) be a non-negative integer function defined on the set of positive integers. Let p(t) be a functions, whose values are between 0 and 1, defined on the set of positive integers. An evolving random simplicial complex is a discrete time process S(t) where t assumes integer values, and for every value t, the expression S(t) is a random simplicial complex obtained from S(t-1) by adding n(t) new points, and creating new simplexes with probability p(t), according to the rules of definition 1.

Definitions 1 and 2, together with a series of variations which we will discuss elsewhere, are the foundation of a theory of simulation for social aggregates. Clearly, definition 1 provides the stage for the description of the possible outcomes when a group of individuals have opportunities for the creation of relationships. However, it presents a stable system, which does not account for arrival of new individuals, nor does it include the possibility of relationships dissolving through time. Therefore, any realistic model will have to account for a variety of different options. For example, instead of having a probability p which does not depend on the dimension of the simplexes which are being added, one may imagine that such a value may be an increasing function as the dimension grows. A second option which must be considered has to do with the possibility to delete existing connections, as time grows. So, instead of a single probability p, or p(t), we need to envision two of them; one describes the likelihood that new connections or simplexes are created, while the other describes the likelihood that an existing connection is removed. Rules will have to regulate the implications of the removal of some connections (for example if we remove [A,B] from [A,B,C], then we lose the 2-simplex, and not just the 1-simplex [A,B]. This rule is a consequence of the fact that any modification should still maintain the structure of simplicial complex.

It should be clear from the above that the simple-minded model and its variants that we have sketched can be enriched as needed. For instance, just like recent years have witnessed the shift from random graph to other forms of networks that accounts for hubs, we envision random simplicial complexes where certain nodes or groups of nodes tend to be hubs for higher-dimensional aggregates. More complicated simplicial models can be generated by assuming some correlation between dynamics in two consecutive dimensions (for instance, the probability of forming a group between individuals that are already connected to one another should be much higher than the one afforded by a group of individuals meeting at a party for the first time), as well as the possibility of using conditional probabilities. While we intend to pursue a systematic study of these possibilities in a subsequent work (Anonymous, 2006), we would like to close this section with a description of one additional construct, which may prove useful in the modeling we have proposed.

To begin with, we note that unless we have some information on the structure which we want to model it will be impossible (or at least very hard) to create a reasonable model. In the examples above, for instance, we have assumed that all nodes are equivalent, and that all possible edges (and faces of higher dimensions) are equivalent, in the sense that the probability distribution is uniform. But of course this is a rather coarse approximation of reality. In real life, we know that some people are more likely than others to create connections. In fact, the entire Small World theory discussed in (Watts, 1999b) is predicated on the existence of 'hubs' (i.e. of nodes that have a much larger connectivity with other nodes). It is the existence of these hubs that allows the theory of the six degrees of separation (Watts, 2004) to be formulated.

In our setting, *these hubs must exist at higher dimensions as well*. So, some nodes are more likely to be connected to other nodes, and other nodes may have higher likelihood to be parts of faces or even n-faces. This is clearly the characteristic of a leader (if we are thinking in terms of social networks) or of an opinion leader. Thus, we can define a notion of *simplicial leadership rank*. To do so, we need a small modification of the definition of random simplicial complex. Recall, a random simplicial complex is a set of vertices and a fixed, or constant, probability, which describes the likelihood that edges connect given vertices, and that higher order simplexes exist. We now define the notion of hierarchical random simplicial complex as follows.

*Definition 3*. A hierarchical random simplicial complex is a triple H=(n,N,p), where n and N are positive integers as before, but p is an N-dimensional vector of probabilities (real numbers between 0 and 1). The interpretation is that for every pair of vertices $P_i$ and $P_j$, the value p[(i,j)] describes the probability that $P_i$ and $P_j$ are connected by an edge. Similarly, for every triple of vertices $P_i$, $P_j$, and $P_k$, the value p[(i,j,k)] describes the probability that, if the three vertices are connected by three edges, the complex actually contains also the triangle $[P_i,P_j,P_k]$. Thus, we see that p is defined on any subset of length N<=n of the set of the n vertices. We also see that p is invariant under permutations (so that p[(i,j)]=p[(j,i)], as well as p[(i,j,k)]=p[(i,k,j)], etc.).

In order to simplify the next definition (which is complicated only from a formal point of view, but actually quite simple from an intuitive point of view), we will assume that if V and W are two vertices in a complex, we will denote by T(V,W,k) the set of all collections of k vertices in the same complex, which do not contain either V or W.

*Definition 4.* A vertex V in a random simplicial complex is said to be a simplicial leader of dimension k (k any integer between 1 and N) and order t (t a number between 0 and 100), or simply a (k,t)-hub, if for any vertex W different from V, the number of elements T in T(V,W,k) for which p[(V,T)] > p[(W,T)] is bigger than t% of the number of elements in T(V,W,k). Obviously, this definition simply means that (if t is close to one, at least), the vertex V has larger probabilities than most of the other vertices, as far as k-simplexes are concerned.

As we said before, specific information is needed if we are to assign appropriate leadership indices to different nodes. In (Sparks-Bethea, 2002) for example, the author shows how to apply social sciences methodology to determine or at least approximate possible indices to different components of family triads. Even though (Sparks-Bethea, 2002) predates this paper by several years, we think that many of its conclusions can be neatly framed in this more mathematically rigorous model, and the ideas suggested there can be useful to create reasonable models of family interactions.

It is also worth noticing that one could take issue with our specific definition of simplicial leader. One may point out that such a definition reflects a sort of democracy in the society described by our model. Indeed, we define a leader as somebody who is involved in a majority of the social aggregations in the society. Different structures (for example the structure of a terrorist organization such as Al Qaeda) may need to be modeled differently. A leader such as Osama Bin Laden may not be involved in large groups where ideas are discussed. Rather, he is probably at the top of a one-dimensional pyramid which connects him to a large number of cells. Once he issues a declaration, he is able to influence a large number of cells, who automatically react to his orders, even though none of the members of these cells can be said to be part of the leadership group. In other words, defining and operationalizing the term leader will likely result in differential representations depending upon the social structure within which the leader is embedded.

## Some computational remarks

This last section contains an initial attempt to sketch a computational approach to these ideas, as well as describe its possible application in concrete social analysis. The notes below are the bare bones of our approach, and serve as an indication of future lines of research.

The reader may recall from Section 3 that, assigned a social group (i.e. a simplicial complex), its Betti numbers gauge its level of connectivity in the various dimensions. A Betti number of dimension n measures how many n-dimensional holes are contained in the group as a whole. What kind of information does this number convey to the social analyst? Let us assume the complex has third Betti number equal 1. This mean that somewhere in the complex a three dimensional subcomplex is missing (i.e. a three dimensional object assembled from simplices of dimension at most three), though all of its faces are already there. Depending on the context, the analyst may conclude that the complex is weak in terms of triadic relations. He or she may also anticipate, depending of course on the statistics governing the complex, that such group will "complete" itself by "filling the hole" at some later stage. Strategic war analysts may be interested in the current status of the complex in an attempt to disconnect it in some critical areas (i.e. disrupt its cohesiveness). Unlike networks, general simplicial complexes can be connected or disconnected in higher dimensions than 1, an extremely important fact for both strategic warfare and social interactions. For instance, terrorist organizations may want to destroy cohesiveness in such social structures, whereas in family relationships, the goal may be to selectively increase cohesiveness.

Many more techniques of social engineering can be inferred by these calculations. We should point out that Betti numbers are in a sense the easiest "homological invariants" of simplicial complexes. More refined tools from homological algebra and homotopy theory are available, representing more accurate gauges of the complex.

But, how can we turn these principles into actual computer calculations? The good news is that homological algebra can be tackled computationally. Indeed, the original motivation for combinatorial objects such as simplicial complexes was precisely their intrinsic computational nature. A considerable amount of work has been done to automatize homological computations (Kaczynski et al., 2004).

The Computational Homology Project (CHOMP, n.d.) has pioneered a software library to efficiently perform homological calculations. Software tools such as CHOMP will have to be enriched to include the capability of generating random simplicial complexes and track down their evolution. We envision the creation of an integrated software environment that will also enable the analyst to interactively manipulate a social complex and visualize parts of it by selecting sub-complexes and set their properties.

The fundamental, ground-level, paradigm of programming, as in (Knuth, 1997) is that programs are nothing but a set of data plus a set of instructions. High level programming requires ways to structure data according to specific methods. Up to this point, most data structures are either relational (such as in data bases) or trees, linked lists, queues, or even simple arrays and vectors. Our idea is to use abstract simplicial complexes as intrinsically hierarchical and dynamic data structures. Abstract simplicial complexes can be thought of as "realizations" of the simplicial paradigm in different settings. In fact, mathematicians would say that they are functors from the template simplicial category $\Delta$, to the category of the data which we want to structure. Data assembled in this way allow the consideration of a hierarchical structure, which is natural in the setting of social interactions (where higher order interactions find their foundations in lower level interactions), as well as the modeling of their rich dynamics (creation, destruction, and evolution of cells within larger social groups). Our forthcoming (Mannucci, Mircea, Sparks, Struppa, 2006a) article will develop these ideas in detail.

Before we conclude, let us also offer a few suggestions for possible measures one may want to look when modeling social aggregates through simplex theory. Let us assume we are looking at an organization such as the membership of Facebook.com; a way to understand the nature of its members is through a monitoring of the usage of the software, and the number of messages exchanged. Similarly, when studying a terrorist organization, we may be able to intercept emails and phone messages. Their numbers will offer us an insight on the structure of the simplicial complex describing the organization. We can use those data to analyze and make predictions on the leadership structure of the organization. The velocity of diffusion is another important element that can be monitored to gain information on a partially unknown simplicial complex. Suppose, for example, that we are able to infiltrate two small cells in a large terrorist organization. We have a glimpse of what the structure looks like in those cells, but we have no information as to how the cells are connected. We could, however, send a message through one of our informants, and see how long it takes for the message to reach the other informant, who is working at a completely different physical location. If repeated several times, with different entry points, and different exit points, one may use this information to infer properties of the entire organization. Although such information may be difficult to obtain, we can still input some data to see how long it takes the organization to get to a "chatter" level or state and subsequently how long it would take for such "chatter" to dissipate. Such analyses would reveal to what extent the individuals in the organizations are connected to each other or not.

In conclusion, we have offered a mathematical model, which describes social aggregations, provides a way to mathematically classify them, and finally is flexible enough to allow for computational experiments to study the natural evolution of such social aggregations. The model we have proposed can be modified to accommodate different needs on the part of the social scientist, and particularly contributes to a more precise understanding of the underlying functions of interpersonal and intergroup communication.

Social Aggregation    29

Social Aggregation    30
Sparks-Bethea, L. (2002). The impact of an older adult parent on communicative satisfaction and dyadic adjustment in the long-term marital relationship: Adult children and spouses' retrospective accounts. *Journal of Applied Communication Research, 30* (2), 107-125.

Watts, D.J. (1999a).   Networks, dynamics and the small world phenomenon, *American Journal of Sociology, 105(2),* 493-527.

Watts, D.J. (1999b). *Small Worlds: The Dynamics of Networks Between Order and Randomness*. Princeton: Princeton University Press.

Watts, D.J. (2004).  *Six Degrees: The Science of a Connected Age*. W. W. Norton & Company.

Wheeler L. & Nezlek, J. (1977). Sex differences in social participation. *Journal of Personality and Social Psychology, 35,* 742-754.

Wilmot, W.W. (1987). *Dyadic Communication* (3rd ed.). New York: McGraw-Hill.

Wilmot, W.W. & Sillars, A.L. (1989). Developmental issues in personal relationships. In J.F.Nussbaum (Ed.), Life-Span communication. Normative processes (pp. 120-132). Hillsdale, NJ: Lawrence Erlbaum.
Mirco A. Mannucci

HoloMathics, LLC

mirco@holomathics.com

Lisa Sparks & Daniele C. Struppa

George Mason University

lsparks@gmu.edu

dstruppa@gmu.edu

Note: starting fall 2007, the last two authors will be at Chapman University.

Their new addresses will be:

sparks@chapman.edu

struppa@chapman.edu